\documentclass[11pt]{article}
\usepackage{epsfig}
\usepackage[utf8]{inputenc}
\usepackage[margin=0.25in]{geometry}
\usepackage{color}
\usepackage{amssymb}
\usepackage{amsmath}
\usepackage{hyperref}

\usepackage{caption} 

\begin{document}
\pagestyle{plain}
\newcount\eLiNe\eLiNe=\inputlineno\advance\eLiNe by -1
\title {Number of equidistant neighbors on honeycomb lattice.}
\author{Zbigniew Kozio{\l}\footnote{e-mail: softquake@gmail.com}\\
National Center for Nuclear Research, Materials Research Laboratory,\\
ul. Andrzeja Sołtana 7, 05-400 Otwock-Świerk, Poland}

\maketitle 

\begin{abstract}
A convenient scheme is presented for calculating potential energy of van der Waals interacting bilayer graphene and other similar 2D compounds. It is based on the notion of the existence of two types of local symmetry of carbon atoms ordering, a 3- and 6-fold one. Potential energy of an atom is expressed as a sum of contributions from rings of equidistant atoms on neighboring layer. Methods are described to compute the radius of rings of equidistant atoms and number of atoms they contain. Exact positions of atoms  are found as well, allowing to apply the introduced method in modelling of anisotropic potentials and to be used when twisting between layers is present.
\end{abstract}


\tableofcontents

\section{Van der Waals potential energy of an atom on graphene structure.} 

After the experimental realisation of single graphene layers \cite{geim07} a growing interest arose in artificial 2-dimensional structures. Other newly discovered materials, named \emph{van der Waals heterostructures} \cite{Geim} become the subject of broad research. The reason of that is richness of physical phenomena observed and high potential of their possible applications \cite{Nam}. 
Novel materials include graphene/silicene, graphene/$MoS_2$, silicene/$MoS_2$ systems \cite{Nam}, as well silicene \cite{Houssa} or $MoS_2$ alone \cite{Kvashnin}. 

When two layers of graphene are rotated \cite{Wijk},\cite{Wijk2} Moiré patterns result. They are related to formation of spatial patterns due to singularities in density of electronic states. In bilayer graphene twisted at about 1.06$^{\circ}$ an unconventional superconductivity was found with critical temperature of 1.7K \cite{YuanCao}, \cite{Bistritzer}, \cite{Tarnopolsky}, \cite{Cao2}. The discovery led to formation of a new research direction named \emph{twistronics}, where physical material properties are controlled by rotation of layers along with traditional methods such as chemical doping, electrical and magnetic field or pressure. Coexistence of insulating states mixed with superconductivity has been controlled by electric field and twist angle, offering a chance for theoretical bridging novel materials with high-$T_c$ superconductors \cite{Guinea}, \cite{Codecido}, \cite{Lu}.

For a proper understanding of the role of interlayer van der Waals forces in these 2D materials it is desirable to expand available theoretical concepts, for the use along the existing methods such as for instance these in large scale modelling of huge collections of atoms in molecular dynamics simulations \cite{LAMMPS}. With right computational algorithms there is no difficulty finding the number and positions of neighbors to carbon atom on graphene lattice. An insight into the role of symmetries, possible broader implications may be overlooked however when purely computer solutions are at use.

There are two types of relative ordering of two layers of graphene that are of special importance in any theoretical considerations. These are named AA ordering, when atoms of one layer are on top of atoms of another layer, and AB stacking when some atoms of one layer are located over the centers of hexagon cells of another layer. AB ordering is energetically stable and common in nature while AA order is observed in some rear, special cases only. When there is some disorder in arrangements of layers, or there is a rotation between their lattices, locally both orderings are found.

\begin{figure}[ht]
\centering
\includegraphics[scale=0.5]{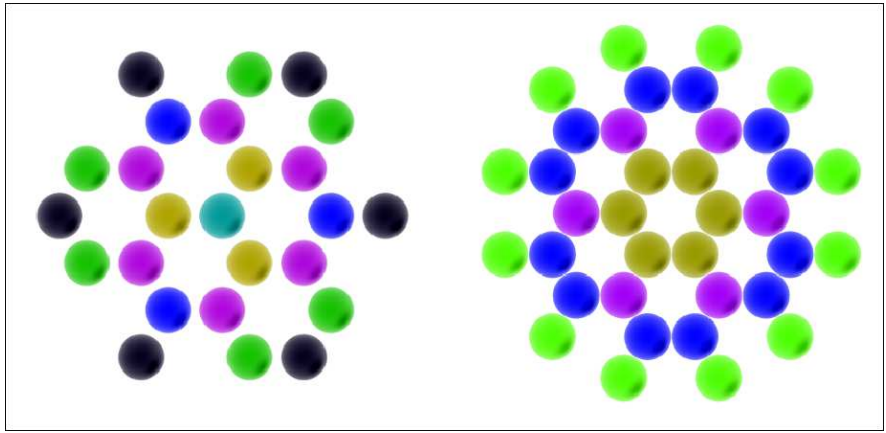}
      \caption{\label{ringsAAAB}
Rings of equidistant atoms on neighboring planes of graphene. On the left is type I ordering, present for AA and AB stacking and type II (right) present in AB stacking, only.
}
\end{figure}

In case of AA stacking we will call the local ordering of atoms on neighboring layers a type I, and for AB stacking - a type II, as shown in Fig. \ref{ringsAAAB} (type I is also present for AB planes while type II can be found for AB structures, only). 

Potential energy $\Phi _{I}(z)$ for an atom in position AA at a distance $z$ from the plane is given by an infinite sum of contributions from equidistant atoms of type $I$:

\begin{equation}\label{MN_02}
\Phi _{I}(z) = \sum _{i=1}^{\infty} N(i) \cdot U\left(\sqrt{(z^2 + (a \cdot \rho_i)^2})\right),
\end{equation}

where the number of neighbors $N(i)$ of equal distance depends on distance projection $\rho(i)$ measured within the plane and $a$ is the unit cell length for in-plane atoms ($a=1.42$ {\AA} in case of graphene). The potential energy between two single atoms, $U$, is any of potentials discussed often in literature (Lennard-Jones, Mie \cite{Mie}, Kolmogorov-Crespi \cite{Kolmogorov}, \cite{Kolmogorov2}, Lebedeva's \cite{Lebedeva}, \cite{Lebedeva2}, as reviewed in \cite{Koziol}). 

In case of AB planes, there is equal number of atoms that are in ordering of type $I$ and type $II$. Equation on energy for type $II$ atoms, $\Phi _{II}(z)$, is similar to Eq. \ref{MN_02} but the number of neighbors at equal distances is different. In case of AB ordering we must take an average of $\Phi _{I}(z)$ and $\Phi _{II}(z)$ as an average energy of an atom. 

A quick convergence was found \cite{Koziol} of potential energy $\Phi$ as a function of the total number of atoms $N$ in a symmetric \emph{molecule} for type I and type II \emph{molecules}: an approximately $1/N^2$ scaling is found (or $1/R$, where $R$ is \emph{molecule} radius), when the rings of equidistant neighbors are added in Eq. \ref{MN_02}.

A convenient scheme may be created of calculating potential energy for different stackings by taking advantage of local symmetries of neighboring atoms. For that we need projection (in plane) of distances between an atom position over the plane together with number of atoms at these distances (i.e. atoms in rings of equidistant atoms, as these in Fig. \ref{ringsAAAB}). For symmetrical atom-atom interaction potentials there is no need for other information. When potentials of KC and Lebedeva are considered, relative orientation of atoms in $x-y$ plane does not play a role either. When an interaction potential depends on orientation between atomic orbitals within $x-y$ plane, like a one introduced recently by Carr et al. \cite{Carr}, exact coordinates of atoms are needed in computation, as provided by the method we propose.

\section{Radius of rings of equidistant atoms and number of their atoms.}

\subsection{Radius of rings.}

Positions $(x,y)$ of carbon atoms in a graphene lattice of \emph{C-C} spacing $a=1.42${\AA} may be 
written as\footnote{This is a convenient convention, in particular when discussing formation of Moiré patterns; One may use other schemes as well, for instance that of dos Santos et al., \cite{lopes}, \cite{lopes2}, or that of Shallcross et al. \cite{Shallcross0}, \cite{Shallcross1}, \cite{Shallcross2}. The basis vectors used by dos Santos et al. are rotated by 60$^{\circ}$ with respect to ours, their order of $(m,n)$ indexes is replaced, their coordinate system is left-handed. After these differences are taken into account any results derived by us reproduce exactly these obtained by dos Santos et al.}:

\begin{equation}
\label{positions}
(x,y) = 	\left( 	m + n/2,	\sqrt{3}\cdot n/2			\right),
\end{equation}

for any pair of integer numbers, $(m,n) \in \mathbb{Z}$, except these when a Boolean function $g(m,n)$ is true:
\begin{equation}
\label{condition}
\begin{split}
 g(m,n) \equiv  (n\mod 3 =0) \wedge (m\mod 3 =1)\\ 
\lor~(n\mod 3 =1) \wedge (m\mod 3 =2)\\ 
\lor~(n\mod 3 =2) \wedge (m \mod 3 =0) = 0 .
\end{split}
\end{equation}

The condition (\ref{condition}) is equivalent to:

\begin{equation}
\label{condition2}
	g(m,n) \equiv \left[(n+1 -(m \mod 3)) \mod 3 = 0 \right].
\end{equation}

Equations \ref{condition} and \ref{condition2} are used for finding indexes $(m,n)$ of positions that are in centers of hexagonal cells not belonging to honeycomb lattice.
The condition that $(m,n)$ belongs to graphene lattice is $G(m,n)$:

\begin{equation}
\label{condition3}
	G(m,n) = !~g(m,n).
\end{equation}


In Equation \ref{positions}, the distance from the center $(0,0)$ of a point defined by indexes $(m,n)$ is given by:

\begin{equation}
\label{rho2}
\rho^2 = m^2+mn+n^2.
\end{equation}

Graphene structure should be considered, in general, as an object possessing a 3-fold rotational symmetry. However, locally, it acquires symmetry 3- or 6-fold, depending on which position is assumed as the center of rotation axis. Also, 6-fold symmetry is preserved when performing any lattice transformations that do not take into account lattice \emph{missing} points described by Eq. \ref{condition3}. 

\begin{figure}[ht]
\centering
\includegraphics[scale=1.3]{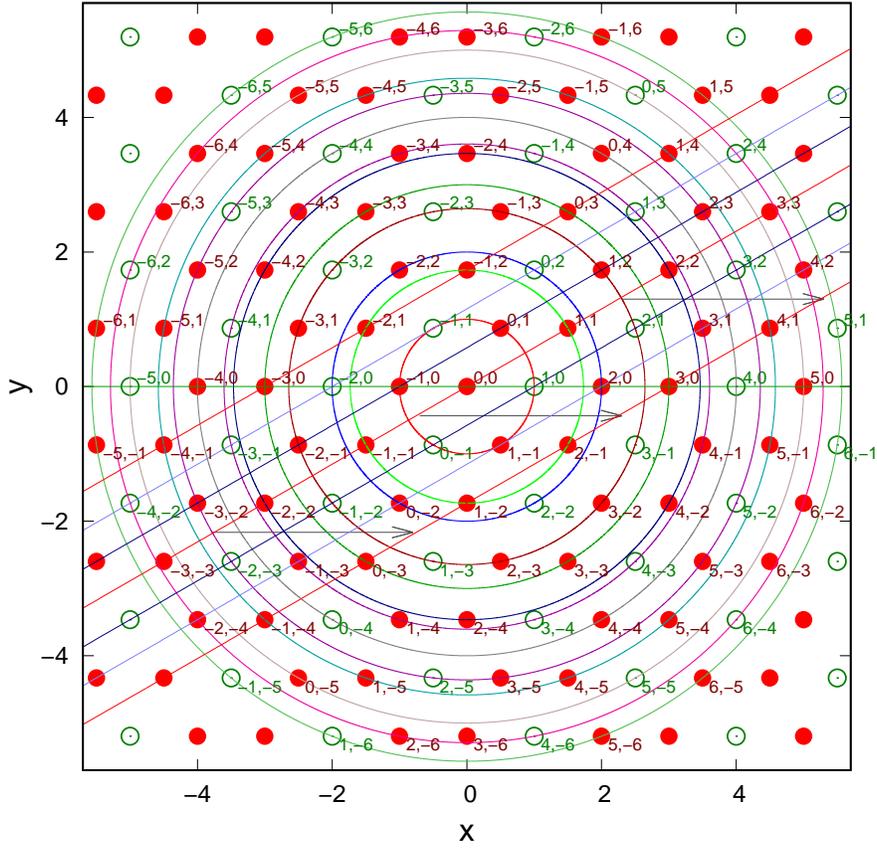}
      \caption{\label{close_to_line_02A}
Graphene lattice (red full circles) created by using Eq. \ref{positions}. Green empty circles denote positions fulfilling Eqs \ref{condition}--\ref{condition3} and they do not belong to graphene lattice. They however fulfill all symmetry properties imposed on Eqs \ref{positions} and \ref{rho2}. Indexes $(m,n)$ are displayed along atoms. Circles centered at $(0,0)$ pass through atoms belonging to rings of equidistant atoms. This lattice, with an atom located in rotational center has the symmetry properties of type I ordering defined in Fig. \ref{ringsAAAB}: a 3-fold rotational symmetry with a broken parity symmetry (a lack of mirror reflection symmetry, e.g. along $0Y$ axis). Lines and arrows are used for explanation of parity symmetry of some atoms.
}
\end{figure}

\subsection{Number of atoms in rings of equidistant atoms}

We observe that the distance expressed by Eq. \ref{rho2} has the following properties:

If a pair $(m,n)$ fulfills one of conditions given by Eq. \ref{condition} than one needs to check if other pairs $(m',n')$ fulfill Eq. \ref{condition}, such that:

\begin{equation}
\label{condition_mn1}
m'=n, n'=m,
\end{equation}

since $m^2+mn+n^2 = n^2+nm+m^2$.

Also we find that the following pair needs to be checked for conditions (\ref{condition}):

\begin{equation}
\label{condition_mn2}
m'=-m, n'=m+n,
\end{equation}

since $m^2+mn+n^2 = (-m)^2-m(m+n)+(m+n)^2$.

We observe also that replacing both signs of $(m,n)$ at the same time also leads to the same distance in Eq. \ref{rho2}.

\begin{equation}
\label{condition_mn3}
m'=-m,~n'=-n,
\end{equation}

Let us define an identity transformation matrix\footnote{Matrices are in capital bold Latin symbols; Sets of matrices are in capital Fraktur style.}, $\mathbf{M}_0$:

\begin{equation}
\mathbf{M}_0
\times
\left[\begin{array}{c}
m\\
n
\end{array}\right]
=
\left[\begin{array}{cc}
1 & 0\\
0 & 1
\end{array}
\right]
\left[\begin{array}{c}
m\\
n
\end{array}\right]
=
\left[\begin{array}{c}
m\\
n
\end{array}\right].
\label{eq:T0}
\end{equation}

The above rules (\ref{condition_mn1},\ref{condition_mn2},\ref{condition_mn3}) can be expressed, respectively (\ref{eq:T0},\ref{eq:T1},\ref{eq:T2}), in the following way, as matrix transformations of $(m,n)$ vector, 

\begin{equation}
\mathbf{M}_1
\times
\left[\begin{array}{c}
m\\
n
\end{array}\right]
=
\left[\begin{array}{cc}
0 & 1\\
1 & 0
\end{array}
\right]
\left[\begin{array}{c}
m\\
n
\end{array}\right]
=
\left[\begin{array}{c}
n\\
m
\end{array}\right].
\label{eq:T1}
\end{equation}

\begin{equation}
\mathbf{M}_2
\times
\left[\begin{array}{c}
m\\
n
\end{array}\right]
=
\left[\begin{array}{cc}
-1 & 0\\
1 & 1
\end{array}
\right]
\left[\begin{array}{c}
m\\
n
\end{array}\right]
=
\left[\begin{array}{c}
-m\\
m+n
\end{array}\right].
\label{eq:T2}
\end{equation}

\begin{equation}
\mathbf{M}_3
\times
\left[\begin{array}{c}
m\\
n
\end{array}\right]
=
\left[\begin{array}{cc}
-1 & 0\\
0 & -1
\end{array}
\right]
\left[\begin{array}{c}
m\\
n
\end{array}\right]
=
\left[\begin{array}{c}
-m\\
-n
\end{array}\right].
\label{eq:T3}
\end{equation}

The above rules may be combined. One finds that $\mathbf{M}_3$ is commutative with $\mathbf{M}_1$ and $\mathbf{M}_2$ while the last two, $\mathbf{M}_1$ and $\mathbf{M}_2$, do not commutate together. Hence, we may generate additional unique transformations. By defining $\mathbf{M}_{ji}$ as $\mathbf{M}_j \times \mathbf{M}_i$, we find: 

\begin{equation}
\mathbf{M}_{21}
\times
\left[\begin{array}{c}
m\\
n
\end{array}\right]
=
\left[\begin{array}{cc}
0 & -1\\
1 & 1
\end{array}
\right]
\left[\begin{array}{c}
m\\
n
\end{array}\right]
=
\left[\begin{array}{c}
-n\\
m+n
\end{array}\right].
\label{eq:T21}
\end{equation}

\begin{equation}
\mathbf{M}_{12}
\times
\left[\begin{array}{c}
m\\
n
\end{array}\right]
=
\left[\begin{array}{cc}
1 & 1\\
-1 & 0
\end{array}
\right]
\left[\begin{array}{c}
m\\
n
\end{array}\right]
=
\left[\begin{array}{c}
m+n\\
-m
\end{array}\right].
\label{eq:T12}
\end{equation}

\begin{equation}
\mathbf{M}_{31}
\times
\left[\begin{array}{c}
m\\
n
\end{array}\right]
=
\left[\begin{array}{cc}
0 & -1\\
-1 & 0
\end{array}
\right]
\left[\begin{array}{c}
m\\
n
\end{array}\right]
=
\left[\begin{array}{c}
-n\\
-m
\end{array}\right].
\label{eq:T31}
\end{equation}

\begin{equation}
\mathbf{M}_{32}
\times
\left[\begin{array}{c}
m\\
n
\end{array}\right]
=
\left[\begin{array}{cc}
1 & 0\\
-1 & -1
\end{array}
\right]
\left[\begin{array}{c}
m\\
n
\end{array}\right]
=
\left[\begin{array}{c}
m\\
-(m+n)
\end{array}\right].
\label{eq:T32}
\end{equation}

There are also tertiary transformations. Of these, $\mathbf{M}_{312}=-\mathbf{M}_{12}$, and $\mathbf{M}_{321}=-\mathbf{M}_{21}$, and we have two more left:

\begin{equation}
\mathbf{M}_{121}
\times
\left[\begin{array}{c}
m\\
n
\end{array}\right]
=
\left[\begin{array}{cc}
1 & 1\\
0 & -1
\end{array}
\right]
\left[\begin{array}{c}
m\\
n
\end{array}\right]
=
\left[\begin{array}{c}
(m+n)\\
-n
\end{array}\right].
\label{eq:T121}
\end{equation}

\begin{equation}
\mathbf{M}_{212}
\times
\left[\begin{array}{c}
m\\
n
\end{array}\right]
=
\left[\begin{array}{cc}
-1 & -1\\
0 & 1
\end{array}
\right]
\left[\begin{array}{c}
m\\
n
\end{array}\right]
=
\left[\begin{array}{c}
-(m+n)\\
n
\end{array}\right].
\label{eq:T212}
\end{equation}

The set of transformations (Eqs \ref{eq:T0} -- \ref{eq:T212}) is complete, in the sense that any transformation of these within that set of any other transformation from within there leads to result that is within the same set of transformations.

\begin{table} 
\begin{center} 
\captionof{table}{Summary of transformation properties of matrices $\mathbf{M}_i$.}
\label{table_MN_summary}
\begin{tabular}{c|c|c|c}
\hline
$\mathbf{M}_i$	&	$\mathbf{M}_i\times(m,n)$	&	$\mathbf{M}_i^{-1}$ & $\mathbf{M}_i^{-1}\times(m,n)$\\
\hline
$\mathbf{M}_0$	&	(m,n)	&	$\mathbf{M}_0$	&	(m,n)	\\
$\mathbf{M}_1$	&	(n,m)	&   $\mathbf{M}_1$	&	(n,m)	\\
$\mathbf{M}_2$	&	(-m,m+n)&   $\mathbf{M}_2$	&	(-m,m+n)\\
$\mathbf{M}_3$	&	(-m,-n) &   $\mathbf{M}_3$	&	(-m,-n)\\
$\mathbf{M}_{21}$	&	(-n,m+n)	& $\mathbf{M}_{12}$	&	(m+n,-m)\\
$\mathbf{M}_{12}$	&	(m+n,-m)	& $\mathbf{M}_{21}$	&	(-n,m+n)\\
$\mathbf{M}_{31}$	&	(-n,-m)		& $\mathbf{M}_{31}$	&	(-n,-m)\\
$\mathbf{M}_{32}$	&	(m,-(m+n))	& $\mathbf{M}_{32}$	&	(m,-(m+n))\\
$\mathbf{M}_{121}$	&	((m+n),-n)  & $\mathbf{M}_{121}$	&	((m+n),-n)\\
$\mathbf{M}_{212}$	&	(-(m+n),n)  & $\mathbf{M}_{212}$	&	(-(m+n),n)\\
$\mathbf{M}_{321}$	&	(n,-(m+n))  & -$\mathbf{M}_{21}$	&	(n,-(m+n))\\
$\mathbf{M}_{312}$	&	(-(m+n),m)  & -$\mathbf{M}_{12}$	&	(-(m+n),m)\\
\hline
\end{tabular}
\end{center} 
\end{table}

Hence, since there are 12 unique transformation matrices, we conclude that there are maximum 12 only possible pairs of $(m,n)$ with the same value of $\rho$, as given by Eq. \ref{rho2} and summarized in Table \ref{table_MN_summary}. We may consider pairs $(m',n')$ that result by applying transformations $\mathbf{M}_i$ to $(m,n)$, as these listed in the second column of Table \ref{table_MN_summary}, as a unique and complete set of elements (group) of a set (group) of transformations we name $\mathfrak{M}^{m,n}$.


Another approach of studying the number of atoms in a ring could go through \emph{classical} rotations.
For counterclockwise 60$^{\circ}$ rotation the transformation matrix $\mathbf{R}_{1}^{xy}$ has the form (we are transforming coordinates of a point $(x,y)$ of positions indexed by $(m,n)$; We distinguish between matrix $\mathbf{R}_{1}^{xy}$ acting in x-y space and $\mathbf{R}_{1}$ acting on indexes $(m,n)$):

\begin{equation}
\left[\begin{array}{c}
x'\\
y'
\end{array}\right]
=
\mathbf{R}_{1}^{xy}
\times
\left[\begin{array}{c}
x\\
y
\end{array}\right]
=
\left[\begin{array}{cc}
cos(\pi/3) & -sin(\pi/3)\\
sin(\pi/3) & cos(\pi/3)
\end{array}
\right]
\left[\begin{array}{c}
x\\
y
\end{array}\right]
=
\left[\begin{array}{cc}
1/2 & -\sqrt{3}/2\\
\sqrt{3}/2 & 1/2
\end{array}
\right]
\left[\begin{array}{c}
m+n/2\\
\sqrt{3}n/2
\end{array}\right]
\label{eq:R10}
\end{equation}

If to write $x'=m'+n'/2, y'=\sqrt{3}n'/2$, one obtains $\mathbf{R}_{1}^{xy}$ in a form acting on indexes 
$(m,n)$:

\begin{equation}
\left[\begin{array}{c}
m'\\
n'
\end{array}\right]
=
\mathbf{R}_{1}
\times
\left[\begin{array}{c}
m\\
n
\end{array}\right]
=
\left[\begin{array}{cc}
0 & -1\\
1 & 1
\end{array}
\right]
=
\left[\begin{array}{c}
-n\\
m+n
\end{array}\right]
\label{eq:R11}
\end{equation}

We find that the matrix $\mathbf{R}_{1}$ of counterclockwise rotation for $\pi/3$ is the same as $\mathbf{M}_{21}$. In a similar way we may find that matrix of clockwise rotation for $\pi/3$, is the same as $\mathbf{M}_{12}$. Obviously, a matrix for 120$^{\circ}$ rotation, $\mathbf{R}_{2}$, may be written as:

\begin{equation}
\mathbf{R}_{2}
=
\left[\begin{array}{cc}
0 & -1\\
1 & 1
\end{array}
\right]
\times
\left[\begin{array}{cc}
0 & -1\\
1 & 1
\end{array}
\right]
=
\left[\begin{array}{cc}
-1 & -1\\
1 & 0
\end{array}
\right]
\label{eq:R2}
\end{equation}

For completeness, let us have explicitly form of rotation matrices for other multiplicities of 60$^{\circ}$: 

\begin{equation}
\mathbf{R}_{3}
=
\left[\begin{array}{cc}
-1 & 0\\
0 & -1
\end{array}
\right],
~~~
\mathbf{R}_{4}
=
\left[\begin{array}{cc}
0 & 1\\
-1 & -1
\end{array}
\right],
~~~
\mathbf{R}_{5}
=
\left[\begin{array}{cc}
1 & 1\\
-1 & 0
\end{array}
\right].
\label{eq:R3_5}
\end{equation}

There is an equivalence between matrices $\mathbf{M}$ and $\mathbf{R}$. Two subsets of transformations $\mathbf{M}_{i}$ may be created:

\begin{equation}
\begin{array}{rl}
\mathfrak{A} & = (\mathbf{M}_{0}, \mathbf{M}_{21}, \mathbf{M}_{312}, \mathbf{M}_{3}, \mathbf{M}_{321}, \mathbf{M}_{12})\\
\mathfrak{B} & = (\mathbf{M}_{1}, \mathbf{M}_{2}, \mathbf{M}_{212}, \mathbf{M}_{31},  \mathbf{M}_{32}, \mathbf{M}_{121}).\\
\end{array}
\label{eq:Msubsets}
\end{equation}

Every next element of subset $\mathfrak{A}$ or $\mathfrak{B}$ in Eq. \ref{eq:Msubsets}, in order as defined there, is related to its previous element by transformation $\mathbf{R}_1$. The last elements transform to first ones. Transition between elements of subsets $\mathbf{A}$ and $\mathbf{B}$ is possible by transformation of indexes $(m,n)\Rightarrow (n,m)$, that is by $\mathbf{M}_{1}$.

\begin{figure}[ht]
\centering
\includegraphics[scale=0.8]{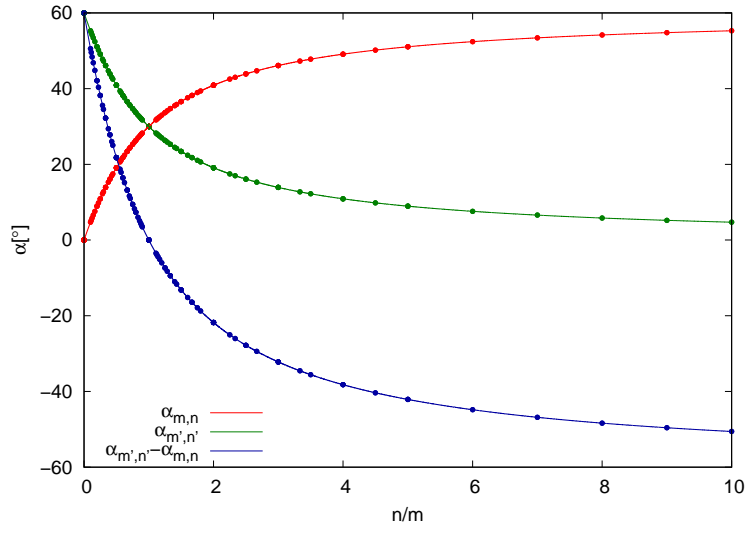}
      \caption{\label{mn_nm_angles_00}
      Change of angles (with respect to $0X$ axis) for positions of points ($m,n)$ and $(m'=n,n'=m)$, as well the difference between them, as a function of $n/m$ ratio. The angles may have discrete values only. Here they are drawn for $n$ and $m$ in the range from 0 to 10. The continuous lines are drawn with equations $\alpha_{m,n}(x)= acos((1+x/2)/\sqrt{1+x+x^2})$ and $\alpha_{m',n'}(x)= acos((x+1/2)/\sqrt{1+x+x^2})$, where $x=n/m$.
}
\end{figure}

Therefore, we find one more way of creating a set of transformations of initial point $(m,n)$ into points of the same distance, by using 60$^{\circ}$ rotations on points $(m,n)$ and $(n,m)$. Figure \ref{mn_nm_angles_00} shows how angles (with respect to $0X$ axis) and difference between them change for points ($m,n)$ and $(m',n')$, where $(m',n')=\mathbf{M}_{1}\times(m,n)$, as a function of $n/m$ ratio.

The parallel lines in Fig. \ref{close_to_line_02A} are drawn for a geometrical explanation of parity symmetry and for finding relation between indexes $(m,n)$ for which parity holds.

The red parallel lines (30$^{\circ}$ slope with respect to $0X$ axis) pass through points belonging to graphene lattice, only. The lines are shifted with respect to each other by vector of length 3 along the  $0X$ axis (as shown by horizontal arrows). Any parallel line shifted for multiplicity of 3 along $0X$ passes through points belonging to graphene lattice, only. For all points lying on the red line passing through $(0,0)$ (all of them have the same $m$ and $n$ indexes) the parity transformation is valid: if $(m,n)$ belongs to graphene then $(m',n') = \mathbf{M_3} \times (m,n)$ belongs to graphene lattice. Hence, for any value of $n$, points with $m=n$ belong to graphene lattice and fulfill the condition of parity symmetry. The right-shifted red line (arrows in Fig. \ref{close_to_line_02A} are shift vectors) passes through points shifted right for 3 by $m$ and it passes points $(3,1), (3,2)$,  etc.

Hence, any points $(m,n)$ fulfilling the condition:

\begin{equation}
\label{eq:parity_condition}
m=n+3k,
\end{equation}

where $k \in \mathbb{Z}$, belong to graphene lattice and have a parity counterpart $(m',n')=(-m,-n)$ belonging to graphene lattice.

Figure \ref{close_to_line_02A} shows lines parallel to these in red, in blue and light blue colors,
 passing through points $m=n+3k+1$ and $m=n+3k+2$. They are also periodic in $m$, with period $3$, however periodicity is present for positive only or negative only values of $m$, i.e. there is a reflection symmetry for these lines with respect to value $m=0$. 
Any point lying on a line $m=n+3k+i$ under transformation $m \leftrightarrow -m$ and $n \leftrightarrow -n$ finds its image on a line $m=n-3k-i$, where $i=1,2$. However, points on blue lines $m=n+3k+1$ do not belong to graphene lattice while all their transform images belong to. In a similar and reverse way, all points on light blue lines, $m=n+3k+2$, belong to graphene lattice while their transformed images do not.

It is convenient now to change the notation. Let us name elements of sets defined by Eq. \ref{eq:Msubsets} by using rotation transformations:

\begin{equation}
\begin{array}{rl}
\mathfrak{T} = (\mathfrak{A}, \mathfrak{B}) = ((\mathbf{R}^{0}, \cdots, \mathbf{R}^{5}),
(\mathbf{R}^{0}\mathbf{M}, \cdots, \mathbf{R}^{5}\mathbf{M})),\\
\end{array}
\label{eq:Msubsets2}
\end{equation}

where $\mathbf{R}^{i}\mathbf{M}$ is $i$ times repeated rotation operation $\mathbf{R}_{1}$ on $\mathbf{M}_1$.

When Eq. \ref{eq:parity_condition} is valid, all 12 available transformations (\ref{eq:Msubsets2}) produce a 12-element ring of 6-fold rotational symmetry. In cases when parity symmetry is broken ($m=n+3k+1$ or $m=n+3k+2$), a 3-fold rotation symmetry holds. That means that from between 6 available rotation matrices $\mathbf{R}^{i}$ in Eq. \ref{eq:Msubsets2} three only transformations are allowed, and from 6 matrices $\mathbf{R}^{i}\mathbf{M}$ as well three only. Again, analysis of positions of lattice points in Fig. \ref{close_to_line_02A} may help in resolving the question on which transformations between graphene atoms are valid: points on $m=n+3k+1$ line are not valid graphene lattice points. However, points that are obtained by applying $\mathbf{R}^{1}$ transformation to them are valid lattice points. Then two other points must be valid as well that are rotated by $120^{\circ}$ and $2\times 120^{\circ}$. Therefore valid transformations in this case are $\mathbf{R}^{1}, \mathbf{R}^{3}, \mathbf{R}^{5}$. 
We see also that point $\mathbf{R}^{0}\mathbf{M}$ (i.e. $\mathbf{M}_{1}$) is a valid lattice point and deduce in a similar way that transformations $\mathbf{R}^{0}\mathbf{M}$, $\mathbf{R}^{2}\mathbf{M}$, $\mathbf{R}^{4}\mathbf{M}$ are all valid. 
By using similar reasoning we find points fulfilling condition $m=n+3k+2$. Let us name these sets as $\mathfrak{C}_1$ and $\mathfrak{C}_2$:

\begin{equation}
\begin{array}{rl}
m=n+3k+1~\Rightarrow~ &\\
\mathfrak{C}_1 = 
(\mathbf{R}^{1},\mathbf{R}^{3},\mathbf{R}^{5},\mathbf{R}^{0}\mathbf{M}$,$\mathbf{R}^{2}\mathbf{M}$,$\mathbf{R}^{4}\mathbf{M})&\\
\end{array}
\label{eq:C1}
\end{equation}

\begin{equation}
\begin{array}{rl}
m=n+3k+2~\Rightarrow~&\\
\mathfrak{C}_2 = (\mathbf{R}^{0},\mathbf{R}^{2},\mathbf{R}^{4},\mathbf{R}^{1}\mathbf{M}$,$\mathbf{R}^{3}\mathbf{M}$,$\mathbf{R}^{5}\mathbf{M})\\
\end{array}
\label{eq:C2}
\end{equation}

There are special cases of $(m,n)$. When $m=n$, there is no difference between points $(m,n)$ and $(n,m$. This is seen also in Fig. \ref{mn_nm_angles_00}, where $\alpha_{m,n}=\alpha_{m',n'}=30^{\circ}$. Therefore, transformations $\mathbf{R}^{i}\mathbf{M}$ in Eq. \ref{eq:Msubsets2} are the same as $\mathbf{R}^{i}$ alone and the set of available 6 transformations is $\mathfrak{A}$ given by Eq. \ref{eq:Msubsets2}.

Another special case is when $n=0$. One may check then (by using for instance simple operations on matrices) that in this case the results of some of transformations of subset $\mathfrak{A}$ are identical to some transformations of subset $\mathfrak{B}$ defined in Eq. \ref{eq:Msubsets2}. For instance, 
$\mathbf{R}^{1} = \mathbf{R}^{0}\mathbf{M}$, $\mathbf{R}^{3} = \mathbf{R}^{2}\mathbf{M}$, etc. 
In result, in that case we should distinguish between cases when $m=3k+1$ and $m=3k+2$. The allowed operations are subsets of these defined by Eqs. \ref{eq:C1} and \ref{eq:C2}: 

\begin{equation}
m=3k+1~~~\Rightarrow~~\mathfrak{C}_{1,n=0} = (\mathbf{R}^{1}, \mathbf{R}^{3}, \mathbf{R}^{5}),
\label{eq:C1_n0}
\end{equation}

\begin{equation}
m=3k+2~~~\Rightarrow~~\mathfrak{C}_{2,n=0} = (\mathbf{R}^{0}, \mathbf{R}^{2}, \mathbf{R}^{4}).
\label{eq:C2_n0}
\end{equation}


\subsection{Multiply pairs of (m,n) for equidistant rings of atoms.}
\label{sub-multiply}

There are points $(m',n')$ which have the same distance $\rho$ as $(m,n)$ (Eq. \ref{rho2}) but do not belong to the set of operations $\mathfrak{M}^{m,n}$. For instance: $(7,0) \notin \mathfrak{M}^{5,3}$ and $(7,7) \notin \mathfrak{M}^{11,2}$.  For $\rho<25$ (Table \ref{table_SAME}), the only other pairs with the same $\rho$ are these $(16,1)$ and $(11,8)$, $(19,0)$ and $(16,5)$, $(13,10)$ and $(17,5)$. Another coincidence of that kind are pairs that are related by a trivial relation: if $(m,n)$ generates an equidistant ring than $(q\cdot m,q\cdot n)$ forms a ring that has radius $q$ times larger, where $q$ is any natural number. In Table \ref{table_SAME} such a case is found for pairs $(21,0)$ and	$(15,9)$, since they are related to $(7,0)$ and $(5,3)$ by common divisor of $3$. At larger values of $(m,n)$ such coincidences are found often. A few of them are triplets of pairs and many are quartets of pairs. When to consider pairs of co-prime numbers only, the first triplet is found for $(80,19)$, the first quartet for $(25,23)$. 

\begin{figure}[ht]
\centering
\begin{tabular}{cc}
\includegraphics[scale=0.75]{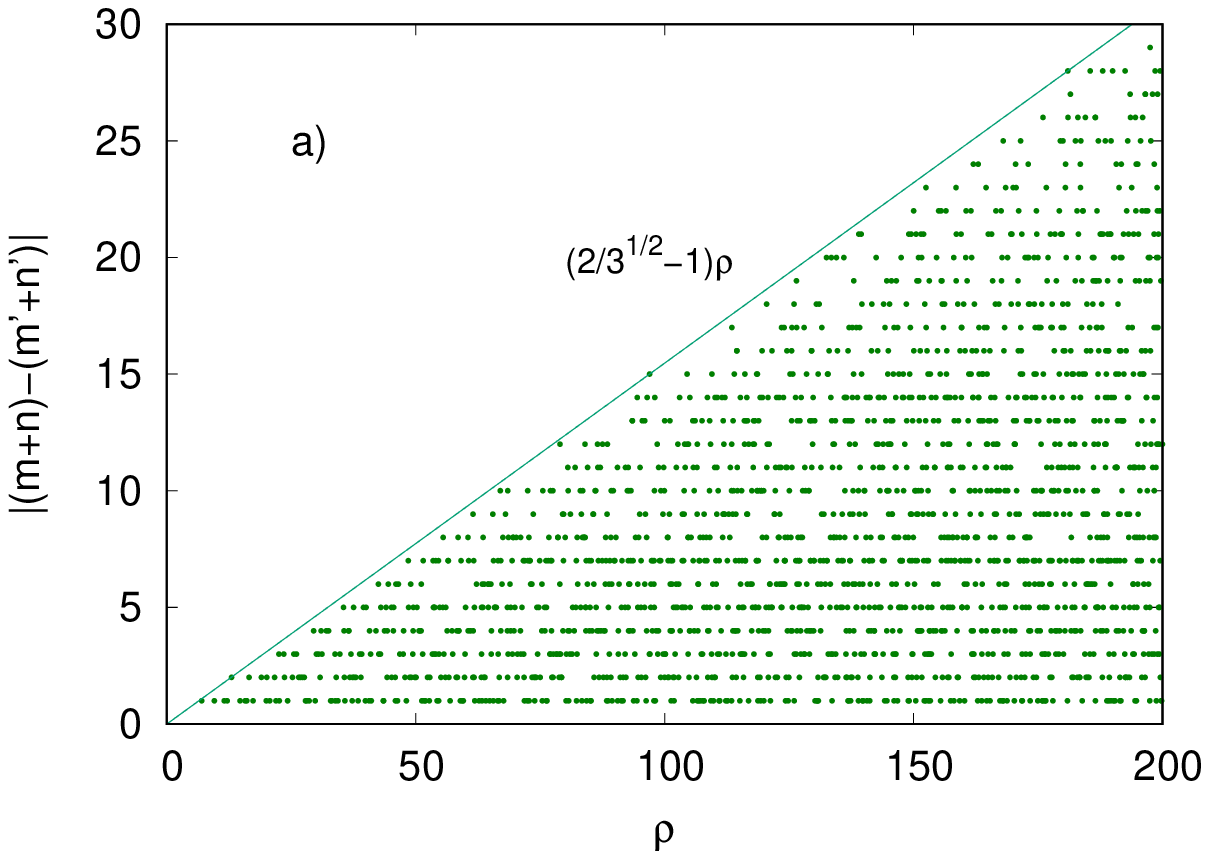}
&
\includegraphics[scale=0.75]{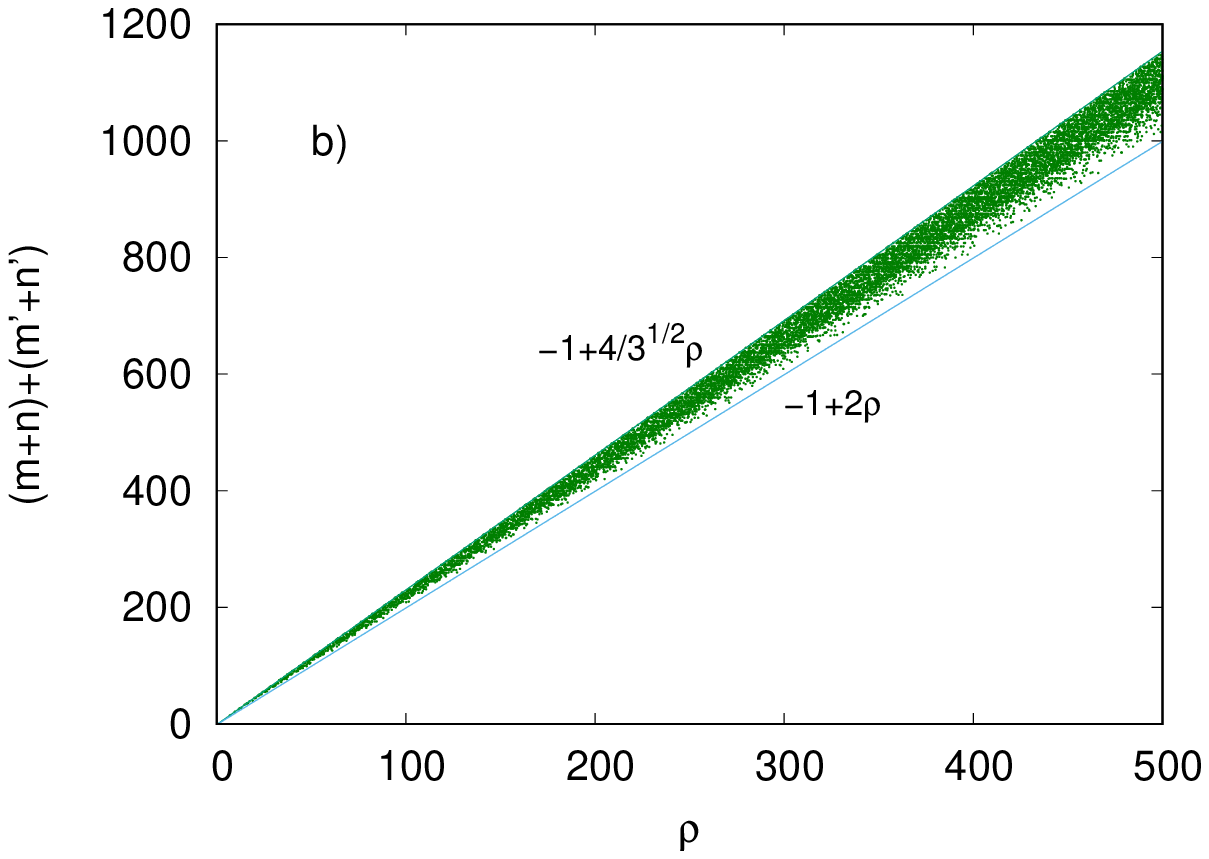}\\
\end{tabular}
\caption{\label{multi_doublets_00} \textbf{a)}: The quantity $|(m+n)-(m'+n')|$ is shown as a function of $\rho$: discrete values of $|(m+n)-(m'+n')|=p$ are found, with integer $p \ge 1$ and $p \le p_{max}$, where $p_{max}$ is integer part of $(2/\sqrt{3}-1)\rho$ and lower than that value. The slope of the limiting straight line can be determined with a very high accuracy and due to that we have confidence that the slope is related to the number $\sqrt{3}$. \textbf{b)}: Another example of restrictions imposed on possible relations between indices of equidistant rings. The slope of upper limiting straight line is $(2/\sqrt{3}-1)\rho$. The lower limiting line is drawn with a slope in a conservative, approximate way. 
}
\end{figure}

\emph{A'posteriori} analysis of the data indicates on existence of a few rules governing the occurrence of multiply pairs of rings with the same values of $\rho(m,n)$, as summarized in Table \ref{Table_restrictions} (Figure \ref{multi_doublets_00} demonstrates some of restrictions on allowed values of $(m',n')$). 

\begin{table} 
\begin{center} 
\captionof{table}{Pairs of $(m,n)$ with the same distance $\rho$ from the ring center, for any $m,n<25$.}
\label{table_SAME}
\begin{tabular}{c|c|c}
\hline
$(m,n)$	&	$(m',n')$	& $\rho^2$\\
\hline
(5,3)	&	(7,0)	&	49\\
(6,5)	&	(9,1)	&	91\\
(7,7)	&	(11,2)	&	147\\
(8,7)	&	(13,0)	&	169\\
(11,8)	&	(16,1)	&	273\\
(16,5)	&	(19,0)	&	361\\
(13,10)	&	(17,5)	&	399\\
(19,3)	&	(17,6)	&	427\\
(21,0)	&	(15,9)	&	441\\ 
(13,12)	&	(20,3)	&	469\\
(19,5)	&	(16,9)	&	481\\
\hline
\end{tabular}
\end{center}
\end{table}

Equation \ref{rho2} may be written in another form:

\begin{equation}
\label{rho20}
\rho^2 = m^2+mn+n^2 = (m+n)^2-mn.
\end{equation}

For pairs of $(m,n)$ and $(m',n')$ of equal-distance $\rho$ we will have:

\begin{equation}
\label{rho21}
(m+n)^2-mn = (m'+n')^2-m'n',
\end{equation}

which may be written as:

\begin{equation}
\label{rho24}
\left[(m+n) +(m'+n')\right] \left[(m+n)-(m'+n')\right] = mn -m'n'.
\end{equation}

In Fig. \ref{multi_doublets_00} a) the quantity $|(m+n)-(m'+n')|$ is displayed as a function of $\rho$: discrete values of $|(m+n)-(m'+n')|=p$ are found, with $p \ge 1$ and $p \le p_{max}$, where $p_{max}$ is integer part of $(2/\sqrt{3}-1)\rho$ and lower than that value. The above condition may be used for finding allowed pairs $(m',n')$ by the method of trial and error, for any reasonably low value of $p$.

\begin{table}  
\begin{center} 
\captionof{table}{Some of conditions on allowed values of pairs $(m',n')$.}
\label{Table_restrictions}
\begin{tabular}{ll}
\hline
1.	&	The sum of indices $m$ and $n$ for rings of the same radius $\rho$ fulfills the condition: $m+n < 2\rho/\sqrt{3}$.\\
2.	&	$0 \le (m-n) \le \rho$;  $0 \le m,n < 1/\sqrt{3} \rho$.\\
3.	&	The sum of indices $m$ and $n$, and $m'$ and $n'$, for rings of the same radius, \\
	&	$(m+n)+(m'+n')$, differs by integer values $p$: $|(m'+n') - (m+n)| = p$, where $p \in [0..p_{max}]$,\\ 
	&	and maximum value of $p$, $p_{max}$, fulfills the condition: $p_{max}< (2/\sqrt{3}-1)\rho$ (Fig. \ref{multi_doublets_00}).\\
4.	&	The sum of multiplicities of indices $m \cdot n$, and $m' \cdot n'$, for different rings of the same radius,\\
	&	is less than 2/3 times $\rho^2$: $m \cdot n + m' \cdot n' < 2/3 \cdot \rho^2$. We have also: \\
	&	$m \cdot n < 1/3 \cdot \rho^2$ and $m' \cdot n' < 1/3 \cdot \rho^2$.\\
\hline
\end{tabular}
\end{center} 
\end{table}


\subsection{Transformations and graphs.}

\begin{figure}[ht]
\centering
\begin{tabular}{cc}
\includegraphics[scale=0.8]{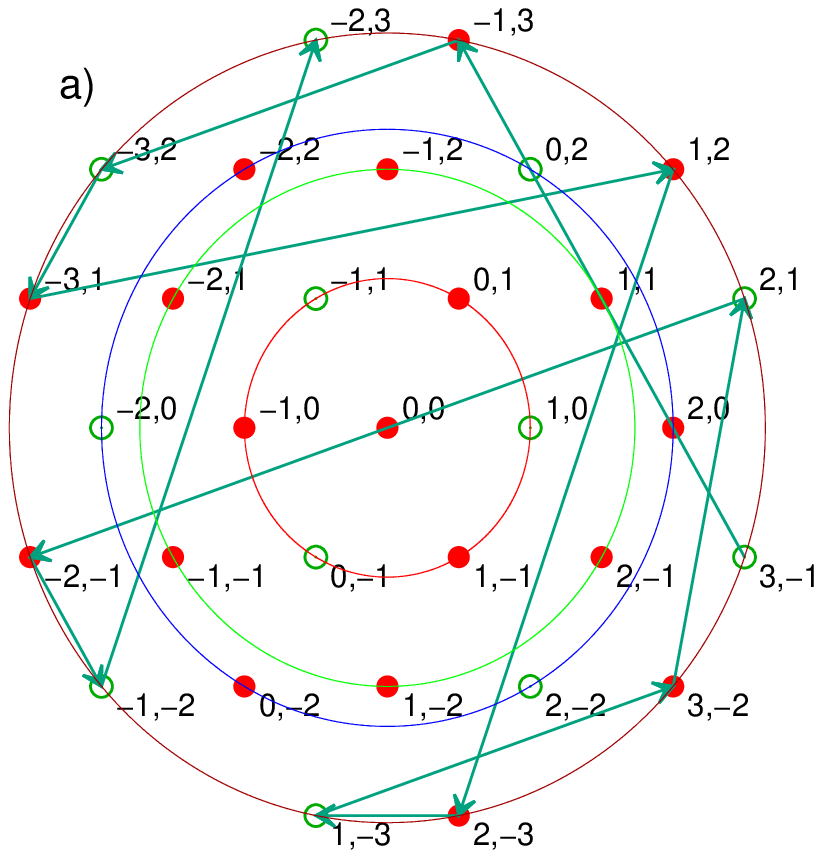}
&
\includegraphics[scale=0.8]{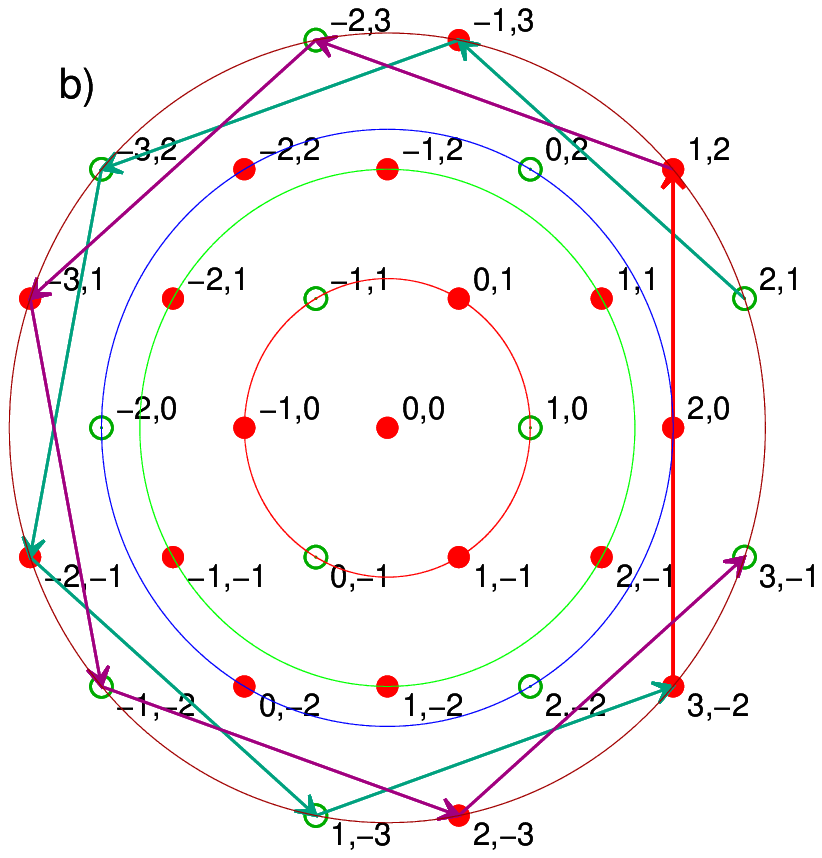}\\
\end{tabular}
\caption{\label{fig:ftransforms0}
Graphene lattice (red circles) and positions not belonging to it (green empty circles). Arrows illustrate transformation paths between points on equidistant rings. In \textbf{a)} the path is from $(3,-1)$ to $(-2,3)$, in \textbf{b)} $(3,-2) \Rightarrow (-1,3)$, with all transformations in the order as listed in Table \ref{table_MN_summary}. Regardless of the starting point of transformation, all points on the ring are visited. In general, however, the transformation paths and number of visited graph vertices depend on the order of transformations. In \textbf{b)} transformations are \emph{ordered}, i.e. each subsequent is a rotation of the previous one by 60$^{\circ}$, except the path between points $(3,-2) \Rightarrow (1,2)$ (red arrow) which uses $(m,n)\Rightarrow (n,m)$ transform between two subsets of 60$^{\circ}$ rotations.
}
\end{figure}

Many aspects of discussed transformation properties may be well illustrated in graphs. Figure \ref{fig:ftransforms0} shows examples of transformation paths between points on equidistant rings. 

In \textbf{a)} the path is from $(3,-1)$ to $(-2,3)$, in the same order as in listing in Table \ref{table_MN_summary}. In \textbf{b)} it is in an ordered manner, as in set $\mathfrak{T}$ in Eq. \ref{eq:Msubsets2}. In general, the transformation paths and number of visited points depend on the order of transformations, points (and paths) may repeat itself. 

\subsection{Type II symmetry.}

In case of type I symmetry, the parity symmetry is broken (there is no symmetry of positions of atoms after reflection along $OY$ axis, or change $x\Rightarrow -x$). In case of type II symmetry, with restored parity symmetry, one more geometrical degree of freedom is available. In result, the number of atoms in rings of the same radius as in case of type I symmetry may become two times larger.

\begin{figure}[ht]
\centering
\begin{tabular}{cc}
\includegraphics[scale=1.0]{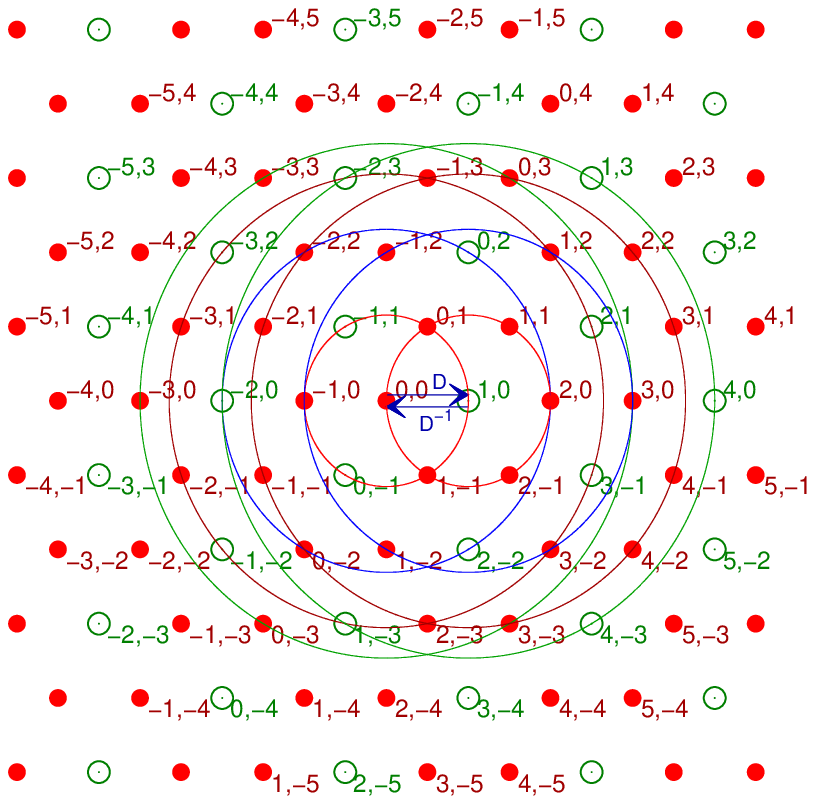}
&
\includegraphics[scale=1.0]{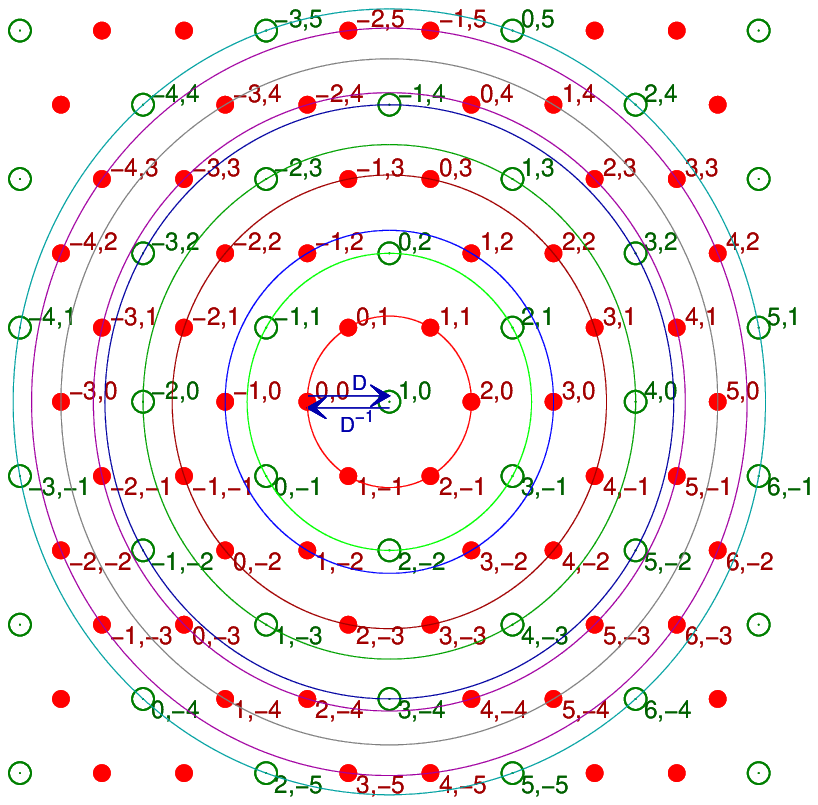}\\
\end{tabular}
\caption{\label{close_to_line_02C} 
By defining displacement operator $\mathbf{D}$, which increases index $m$ by $1$: $\mathbf{D} \times [m,n] = [m+1,n]$, and a reverse one, $\mathbf{D}^{-1} \times [m+1,n] = [m,n]$, any transformation operation on indexes of atoms of type II symmetry may be carried out on by $\mathbf{D}\mathbf{M}_{i}\mathbf{D}^{-1}$. Red full circles belong to graphene lattice while green empty ones do not. The displacement operator $\mathbf{D}$ and its inverse are denoted by $D$ and $D^{-1}$ vectors.
}
\end{figure}

Figure \ref{close_to_line_02C} illustrates that a transition between type I and type II surroundings may be achieved by introducing a displacement operator $\mathbf{D}$ acting on indexes $(m,n)$. It increases $m$ by $1$, while a reverse operator, $\mathbf{D}^{-1}$, decreases $m$ by $1$: 

\begin{equation}
\begin{array}{rl}
\mathbf{D} \times \left[\begin{array}{c}m\\n\end{array}\right] &= \left[\begin{array}{c}m+1\\n\end{array}\right]\\
\mathbf{D}^{-1} \times \left[\begin{array}{c}m+1\\n\end{array}\right] &= \left[\begin{array}{c}m\\n\end{array}\right]\\
\end{array}
\label{eq:D}
\end{equation}

Any transformation operations on indexes of atoms discussed so far for type I symmetry may be carried out on by transformations $\mathbf{D}\mathbf{M}_{i}\mathbf{D}^{-1}$ on atoms of type II symmetry. The only difference in results between I and II symmetries is due to conditions on belonging to graphene lattice (Equations \ref{condition}--\ref{condition3}). These conditions must be checked again in case of operations on atoms of type II. 

Table \ref{matrices_summary} summarizes results on available transformation operations derived for both types of ordering. A column lists there the number of possible graphene atoms in rings, $N$. In parentheses is the number of atoms in type II ordering. For any ring of type II of the same radius as a ring of type I the number of graphene atoms in it is either the same, it is two times larger, or it is none.

\begin{table} 
\begin{center} 
\captionof{table}{
Summary of available transformation operations on co-prime pairs of numbers $(m,n)$ (including $n=0$ and $n=m$: $m \ge n$, $n\ge 0$) preserving the invariant $m^2+mn+n^2$. $N$ is the number of available transformations (for type I ordering; In parentheses - for type II). Here, $k$ is an integer number, $k\ge 0$. Indexes $(m,n)$ refer to type I ordering. 
}
\label{matrices_summary}
\begin{tabular}{l|c|c|c|c}
\hline
Conditions			&	Type I	& Type II	&	N	& Eqs \\
\hline
$m=n+3k, n>0$	&	$(\mathfrak{A},\mathfrak{B})$	& -	&	$12 (-)$ & \ref{eq:Msubsets2}\\
$m=n+3k+1, n>0$	&	$\mathfrak{C}_{1}$	& $(\mathfrak{A},\mathfrak{B})$	&  $6 (12)$ & \ref{eq:C1}\\
$m=n+3k+2, n>0$	&	$\mathfrak{C}_{2}$	& $(\mathfrak{A},\mathfrak{B})$	&  $6 (12)$ & \ref{eq:C2}\\
$m=n$			&	$(\mathbf{R}^{0}, \cdots, \mathbf{R}^{5})$	& -	& $6 (-)$ & \ref{eq:Msubsets2}\\
$m=3k, n=0$		&	$(\mathbf{R}^{0}, \cdots, \mathbf{R}^{5})$	& - & $6 (-)$ & \ref{eq:Msubsets2}\\
$m=3k+1, n=0$	&	$\mathfrak{C}_{1,n=0}$	& $(\mathbf{R}^{0}, \cdots, \mathbf{R}^{5})$	& $3 (6)$ & \ref{eq:C1_n0}\\
$m=3k+2, n=0$	&	$\mathfrak{C}_{2,n=0}$	& $(\mathbf{R}^{0}, \cdots, \mathbf{R}^{5})$	& $3 (6)$ & \ref{eq:C2_n0}\\
\hline
\end{tabular}
\end{center}
\end{table}


\begin{table}  
\begin{center} 
\captionof{table}{In-plane projection of distances between an atom on one graphene plane and nearest neighbors on next plane, for type I of neighbors, as in Fig. 1 (left), found in AA and AB stacking of layers, and for type II ordering, as in Fig. 1 (right). $N$ is a pair of number of equidistant neighbors (in regular font for type I and in bold font for type II ordering). $\rho$ is in units of graphene unit cell value of $1.42$  {\AA}. Pairs $(m, n)$ are example numbers that are used in Equation \ref{positions}.
}
\label{table_AA}
{\small
\begin{tabular}{c|cccccccccccccc}
\hline
$m,n$	&0,0	&1,0	&1,1	&2,0	&2,1	&3,0	&2,2	&3,1	&4,0	&3,2	&4,1	&5,0	&3,3	&4,2\\
$\rho$	&0      &1      &$\sqrt{3}$ &2      &$\sqrt{7}$ 	&3      &$2\sqrt{3}$ 	&$\sqrt{13}$ 	&4      &$\sqrt{18}$	&$\sqrt{21}$ 	&5      &$3\sqrt{3}$ 	&2$\sqrt{7}$ \\
$N$	&1~\textbf{-}	&3~\textbf{6}	&6~\textbf{-}	&3~\textbf{6}	&6~\textbf{12}	&6~\textbf{-}	&6~\textbf{-}	&6~\textbf{12}	&3~\textbf{6}	&6~\textbf{12}	&12~\textbf{-}	&3~\textbf{6}	&6~\textbf{-}	&6~\textbf{12}	\\
\hline
$m,n$	&5,1	&6,0	&4,3	&5,2	&6,1	&4,4	&7,0	&6,2	&7,1	&5,4	&6,3	&8,0	&7,2	&8,1\\
$\rho$	&$\sqrt{31}$ 	&6      &$\sqrt{37}$ 	&$\sqrt{39}$  &$\sqrt{43}$ &$4\sqrt{3}$	&7	 &$2\sqrt{3}$	&$\sqrt{57}$ 	&$\sqrt{61}$ 	&$2\sqrt{7}$ 	&8      &$\sqrt{67}$ 	&$\sqrt{73}$  \\
$N$	&6~\textbf{12}	&6~\textbf{-}	&6~\textbf{12}	&12~\textbf{-}	&6~\textbf{12}	&6~\textbf{-}	&9~\textbf{18}	&6~\textbf{12}	&12~\textbf{-}	&6~\textbf{12}	&12~\textbf{-}	&3~\textbf{6}	&6~\textbf{12}	&6~\textbf{12}	\\
\hline
$m,n$	&5,5	&6,4	&7,3	&9,0	&8,2	&9,1	&7,4	&8,3	&10,0	&9,2	&6,6	&7,5	&10,1	&8,4\\
$\rho$	&$5\sqrt{3}$ 	&$2\sqrt{18}$	&$\sqrt{79}$ 	&9		&$2\sqrt{21}$	&$\sqrt{91}$ 	&$\sqrt{93}$ 	&$\sqrt{97}$ 	&10     &$\sqrt{103}$	&$6\sqrt{3}$	&$\sqrt{109}$	&$\sqrt{111}$	&$4\sqrt{7}$	\\
$N$	&6~\textbf{-}	&6~\textbf{12}	&6~\textbf{12}	&6~\textbf{-}	&12~\textbf{-}	&12~\textbf{24}	&12~\textbf{-}	&6~\textbf{12}	&3~\textbf{6}	&6~\textbf{12}	&6~\textbf{-}	&6~\textbf{12}	&12~\textbf{-}	& 6~\textbf{12}\\
\hline
\end{tabular}
}
\end{center}
\end{table}


\section{Summary and conclusions.}

A scheme is presented for calculating potential energy of van der Waals interacting bilayer graphene and other similar 2D compounds. It is based on the notion of the existence of two types of local symmetry of carbon atoms ordering, a 3- and 6-fold one (type I and type II). Potential energy of an atom is expressed as a sum of contributions from rings of equidistant atoms on neighboring layer. Methods are described to compute  radiuses of rings of equidistant atoms and number of atoms they contain. The methods are based on analysis of transformation properties of equation on atoms arrangement on equidistant rings and on symmetries of underlying crystallographic lattice. Table \ref{matrices_summary} summarizes available transformation operations and number of atoms on rings for both types of orderings for pairs of co-prime numbers $(m,n)$ indexing atoms positions in lattice. Table \ref{table_AA} lists the data obtained for equidistant rings of radius up to about 12 {\AA}.

\begin{figure}[ht]
\centering
\includegraphics[scale=1.0]{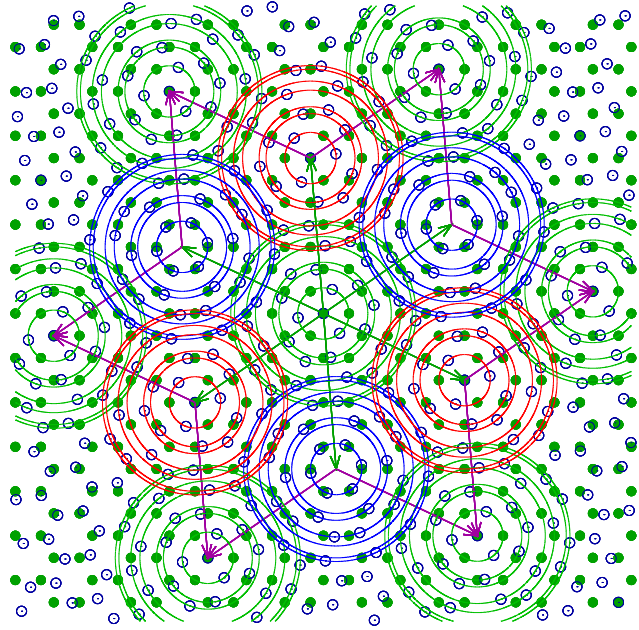}
      \caption{\label{magic_3_1_02}
Moiré pattern for two layers of graphene rotated for \emph{magic} angle 9.430007$^{\circ}$ (\emph{magic} indexes $(3,1)$). Full green circles represent atoms of unrotated layer and blue empty ones of rotated one. Arrows indicate basis vectors forming Moiré superstructure. When to neglect distinction between atoms on different layers than a 6-fold symmetry of Moiré pattern is observed. There are however differences between arrangement of atoms on equidistant rings (blue and green concentric circles) formed around vertices of unit cells, restoring the 3-fold symmetry of the structure. 
}
\end{figure}

Not only the number of neighbors is found but their exact coordinates as well (parameterized by indexes $(m,n)$). That allows to apply the introduced method to modelling anisotropic potentials, when potential energy depends on angles between bonds, and in particular it allows to compute properties of structures formed by twisted bilayers of graphene (Fig. \ref{magic_3_1_02}). For any atom its neighbors may be described with an infinite series of concentric rings. For atoms located at superstructure lattice nodes the centers of rings of atoms from both layers coincide, with arrangement of atoms rotated between layers. When to neglect distinction between atoms on different layers than a 6-fold symmetry of Moiré pattern is observed. There are however differences between arrangement of atoms on equidistant rings (blue and green concentric circles in Fig. \ref{magic_3_1_02}) formed around vertices of unit cells, restoring the 3-fold symmetry of the structure. 

The methods described pave the way towards applications for other 2D structures than graphene alone, for instance for h-BN and graphene/h-BN. Methods of computing strain and stress can be developed, spatially changing chirality is open for investigation. 


\addcontentsline{toc}{section}{\numberline {4} References}

\end{document}